\documentclass[twocolumn]{aastex631}
\usepackage{CJKutf8}
\usepackage{array, booktabs, makecell}

\newcommand\mearth{M$_\oplus$}
\newcommand{\dpn}[3]{$#1\,{\rm #2}_{#3}$}

\received{April, 2021}
\submitjournal{AJ}

\shorttitle{DPs Long-term Evolution}
\shortauthors{Mu\~noz-Guti\'errez et al.}
\graphicspath{{./}{figures/}}

\begin{document}

\title{Long-term Dynamical Stability in the Outer Solar System I: The Regular and Chaotic Evolution of the 34 Largest Trans-Neptunian Objects}

\correspondingauthor{Marco A. Mu\~noz-Guti\'errez}
\email{mmunoz@asiaa.sinica.edu.tw}

\author[0000-0002-0792-4332]{Marco A. Mu\~noz-Guti\'errez}
\affiliation{Institute of Astronomy and Astrophysics, Academia Sinica, 11F of AS/NTU
Astronomy-Mathematics Building, No.1, Sec. 4, Roosevelt Rd., Taipei 10617, Taiwan, R.O.C.}

\author[0000-0001-7042-2207]{Antonio Peimbert}
\affiliation{Instituto de Astronom\'ia, Universidad Nacional Aut\'onoma de 
M\'exico, Apdo. postal 70-264, Ciudad Universitaria, M\'exico}

\author[0000-0003-4077-0985]{Matthew J. Lehner}
\affiliation{Institute of Astronomy and Astrophysics, Academia Sinica, 11F of AS/NTU
Astronomy-Mathematics Building, No.1, Sec. 4, Roosevelt Rd., Taipei 10617, Taiwan, R.O.C.}
\affiliation{Department of Physics and Astronomy, University of Pennsylvania, 209 S. 33rd St., Philadelphia, PA 19104 USA}
\affiliation{Center for Astrophysics \textbar Harvard \& Smithsonian, 60 Garden St., Cambridge, MA 02138 USA}

\author[0000-0001-6491-1901]{Shiang-Yu Wang (\begin{CJK*}{UTF8}{bkai}
    王祥宇\end{CJK*})}
\affiliation{Institute of Astronomy and Astrophysics, Academia Sinica, 11F of AS/NTU
Astronomy-Mathematics Building, No.1, Sec. 4, Roosevelt Rd., Taipei 10617, Taiwan, R.O.C.}


\begin{abstract}

We carried out an extensive analysis of the stability of the outer solar system, making use of the frequency analysis technique over short-term integrations of nearly a hundred thousand test particles, as well as a statistical analysis of 200, 1 Gyr long numerical simulations, which consider the mutual perturbations of the giant planets and the 34 largest trans-Neptunian objects (we have called all 34 objects ``dwarf planets'', DPs, even if probably only the largest of them are true DPs). 

From the frequency analysis we produced statistical diffusion maps for a wide region of the $a$-$e$ phase-space plane; we also present the average diffusion time for orbits as a function of perihelion. We later turned our attention to the 34 DPs making an individualized analysis for each of them and producing a first approximation of their future stability. 

From the 200 distinct realizations of the orbital evolution of the 34 DPs, we classified the sample into three categories including 17 Stable, 11 Unstable, and 6 Resonant objects each; we also found that statistically, 2 objects from the sample will leave the trans-Neptunian region within the next Gyr, most likely being ejected from the solar system, but with a non-negligible probability of going inside the orbit of Neptune, either to collide with a giant planet or even falling to the inner solar system, where our simulations are no longer able to resolve their continuous evolution.

\end{abstract}


\section{Introduction} 
\label{sec:intro}

The outer regions of the solar system, those beyond the orbit of Neptune, are far from empty; instead they contain vast amounts of large and small objects, the remnants of planet formation. Those remnants include millions of small icy planetesimals which constitute the reservoir of visible comets \citep{Dones15,Nesvorny17}, up to dozens of dwarf planet sized objects \citep{Brown04,Schwamb14}, and possibly even planetary sized objects \citep{Trujillo14,Batygin16}. 

In particular, the region immediately after Neptune and up to a few hundred au from the Sun, the trans-Neptunian region, constitutes the reservoir of the short-period comets \citep{Levison97,Duncan04,Volk08}. 

Strictly speaking, the trans-Neptunian region comprises everything beyond Neptune; however it is commonly associated only with the Kuiper belt and its extended region, which includes the detached scattered disk and the extreme trans-Neptunian Objects, but excludes the Oort cloud. Commonly, trans-Neptunian Objects (TNOs) are classified into several sub-populations forming the Kuiper belt, namely: the cold and hot Classical, the Resonant, and the Scattering populations \citep{Gladman08}.

Despite the apparent stability of the Kuiper belt, the mere existence of comets puts in evidence the belt's continuous erosion, as cometary nuclei are driven towards the inner solar system, due to perturbations produced by the giant and dwarf planets, acting on secular time-scales \citep[see for instance][]{Levison97,Volk08,Nesvorny17,Munoz19}. In this regard, the trans-Neptunian region is not absolutely stable, and this lack of stability will, in principle, apply to any object in its dominion, including the dwarf planets themselves.

According to the IAU definition, a dwarf planet of the solar system is a celestial body that: {\it ``(a) is in orbit around the Sun, (b) has sufficient mass for its self-gravity to overcome rigid body forces so that it assumes a hydrostatic equilibrium (nearly round) shape, (c) has not cleared the neighbourhood around its orbit, and (d) is not a satellite.''\footnote{\url{https://www.iau.org/static/resolutions/Resolution_GA26-5-6.pdf}}} The existence of comets makes it clear that, although the orbital cleansing process in the trans-Neptunian region is a slow one, it is nonetheless an active and continuous one which, left to its own devices, could require hundreds of Gyr (or even Tyr) to be completed.

Following the IAU definition, there are currently 5 objects in the solar system recognized as dwarf planets (DPs, hereafter), namely Ceres (in the Asteroid belt), Eris, Pluto, Haumea, and Makemake (in the Kuiper belt). By considering the IAU criteria, a number of other TNOs could potentially be classified as DPs. \citet{Tancredi08} give a list of 18 potential trans-Neptunian DPs, based on physical and dynamical criteria, including the 4 already classified as such \citep[see also and updated list in][]{Tancredi10}. The apparent minimum threshold in radius for TNOs to effectively become a DP, i.e. to attain hydrostatic equilibrium shapes, is around 400 km \citep[see][]{Rambaux17}; this would imply that there are currently about a dozen DPs in the Kuiper belt; on the other end, \citet{Valsecchi09} and later \citet{Margot15} give a criteria to differentiate between planets and DPs, basically the definition of a planet, establishing a clear limit above the size of Eris/Pluto.

We are used to think that members of the solar system (planets, satellites, asteroids, DPs, etc.) which are long-familiar to us, are stable and will remain as such for the age of the solar system and beyond. However, this is not necessarily the case. Actually, even from definition, a DP is an object that has been unable to clean its orbit, thus the constant stirring of smaller objects around could result in destabilizing interactions for both sides. Essentially, in a similar way as the delivery of comets is an example of the slow but continuous process of orbital cleansing in the trans-Neptunian region, larger objects, such as DPs and potential DPs, are prone to experience the same fate and also be removed from the region within the age of the system \citep[though likely at a significantly lower rate than comets; nonetheless, within a much more massive disk this rate should be markedly larger, see e.g.][]{Silsbee19}.

Previous works addressing the stability of the outer solar system have focused mainly on the delivery of comets, thus typically considering only massless particles in numerical simulations \citep[e.g.][]{ Holman93,Duncan95,Robutel01}. In this work, we intend to complement and shed light on the long-term evolution of larger objects, those that in case of being driven inwards, would become a much more spectacular phenomena than normal comets. We build up on our simulations from \citet{Munoz19} to explore the fate of the 34 largest TNOs (including the 4 known DPs). We found that out of the 10 more massive TNOs in our sample, four turn out to be unstable or highly unstable (namely Haumea, \dpn{2007}{OR}{10}, \dpn{2002}{MS}{4}, and Orcus); in our view, this fact alone highlights the importance of considering the interaction among DPs to fairly account for the secular evolution of the outer solar system in general. 


\section{Our sample: The 34 Largest TNOs}

In a previous paper \citep{Munoz19}, we compiled a set of the 34 brightest TNOs known then (early 2019). Since brightness correlates with size and thus with mass, this represents a set of 34 of the most massive TNOs (with some bias for objects closer to the Sun). All four confirmed trans-Neptunian DPs are in this set; probably at least one more is also clearly a DP; regardless, we call DPs indistinctly all the members of this set. In that work we explored the dynamical influence that these objects, a greatly neglected component of the outer solar system, have on the orbital evolution of light TNOs. We specifically focused on the injection rate of cometary nuclei from the Kuiper belt into the inner solar system; we followed the evolution of the light TNOs starting from their interaction with the DPs, continuing with their interaction with Neptune's resonances, a strong interaction with Neptune itself, and up to their end stage as Jupiter Family Comets (JFCs). 

In this work we use the same simulations we used in the previous paper, but we focus our attention, not on the evolution of the test particles perturbed by the DPs, but on the long-term evolution of the DPs themselves. The data for all the 34 members of our DP sample (masses and barycentric orbital elements) are listed in Table \ref{tab:TNOdata1} \citep[here we only recall the basic aspects of the data compilation; for details the reader is referred to the Appendix in ][]{Munoz19}.

Due to their remoteness, physical properties of TNOs are difficult to constrain. However, out of the 34 TNOs in our sample, 13 are known with enough detail to determine their size, mass, and density. Another 14 objects possess a confidently measured radius and albedo; in order to determine their mass we need to know their densities; we used a formula derived in \citet{Munoz19} to assign a density (this formula includes a systematic and a random component) and thus derive the masses for these objects. Finally, for 7 objects only their absolute magnitudes were known, for them we used formulas from \citeauthor{Munoz19} to assign plausible densities and albedos in order to assign plausible radii and masses (both formulas include a systematic and a random component), in order to complete the required input data for our numerical simulations.

Since we ran our simulations, and the publication of our previous paper \citep{Munoz19}, there have been updates to the observed physical parameters of 2002 TC$_{302}$ \citep{Ortiz20}. The absolute magnitude is found to be slightly dimmer (4.32 vs. the old 3.90); also the size was found to be slightly smaller (250 km vs.our assumed 292 km); therefore, the albedo is found to be slightly larger (0.147 vs. the old 0.115); the new size also affects the expected density, which is approximately 15\% smaller than our previous estimate. Overall this would imply a mass of $0.0128 \times 10^{-3}$ \mearth\ instead of the $0.0239 \times 10^{-3}$ \mearth\ that we used (\citeauthor{Ortiz20} prefer a slightly lower density and thus a slightly lower mass of $0.0088 \times 10^{-3}$ \mearth). This difference is not statistically significant as to demand new simulations.
 
There have also been new determinations of the size of Haumea \citep{Ortiz20b}; however, since there are independent determinations of Haumea's mass, and such value has not changed, the new radius does not affect our simulations.

\begin{deluxetable*}{lDcDDDD}
\tablecaption{Main orbital parameters of all 34 DPs}
\label{tab:TNOdata1}
\tablewidth{0pt}
\tablehead{
\colhead{Object} & \multicolumn{3}{c}{Mass ($\times 10^{-3}$ \mearth)} &\multicolumn{2}{c}{a (au)} & \multicolumn{2}{c}{e} & \multicolumn{2}{c}{Inc} & \multicolumn{2}{c}{q (au)} 
}
\decimals
\startdata
Eris               & 2.7956 & Measured  &  67.872 & 0.438 & 43.993 & 38.163 \\
Pluto              & 2.4467 & Measured  &  39.560 & 0.250 & 17.141 & 29.673 \\
Haumea             & 0.6706 & Measured  &  43.144 & 0.194 & 28.205 & 34.772 \\
\dpn{2007}{OR}{10}  & 0.6110 & Measured  &  67.108 & 0.503 & 30.803 & 33.360 \\
Makemake           & 0.5531 & Measured  &  45.542 & 0.159 & 29.002 & 38.286 \\
Quaoar             & 0.2343 & Measured  &  43.393 & 0.037 &  7.991 & 41.773 \\
\dpn{2002}{MS}{4}   & 0.1369 & Estimated &  41.784 & 0.144 & 17.693 & 35.756 \\
Sedna              & 0.1254 & Estimated & 515.069 & 0.852 & 11.929 & 76.190 \\
Orcus              & 0.1073 & Measured  &  39.313 & 0.223 & 20.568 & 30.553 \\
\dpn{2014}{EZ}{51}  & 0.1012 & Estimated &  52.171 & 0.227 & 10.272 & 40.311 \\
\dpn{2010}{JO}{179} & 0.0635 & Estimated &  78.485 & 0.499 & 32.043 & 39.356 \\
\dpn{2002}{AW}{197} & 0.0606 & Estimated &  47.263 & 0.129 & 24.382 & 41.151 \\
\dpn{2015}{KH}{162} & 0.0594 & Estimated &  61.877 & 0.332 & 28.860 & 41.339 \\
Varda              & 0.0446 & Measured  &  45.804 & 0.142 & 21.511 & 39.281 \\
\dpn{2007}{UK}{126} & 0.0415 & Measured  &  73.690 & 0.490 & 23.357 & 37.559 \\
\dpn{2013}{FY}{27}  & 0.0391 & Estimated &  58.808 & 0.394 & 33.120 & 35.641 \\
\dpn{2003}{AZ}{84}  & 0.0349 & Measured  &  39.512 & 0.178 & 13.565 & 32.472 \\
\dpn{2015}{RR}{245} & 0.0331 & Estimated &  81.992 & 0.586 &  7.552 & 33.946 \\
\dpn{2003}{OP}{32}  & 0.0288 & Estimated &  43.247 & 0.106 & 27.161 & 38.654 \\
\dpn{2014}{UZ}{224} & 0.0266 & Estimated & 109.450 & 0.650 & 26.785 & 38.312 \\
Ixion              & 0.0263 & Estimated &  39.550 & 0.244 & 19.631 & 29.898 \\
Varuna	           & 0.0259 & Measured  &  43.005 & 0.053 & 17.175 & 40.709 \\
\dpn{2005}{RN}{43}  & 0.0255 & Estimated &  41.569 & 0.024 & 19.272 & 40.578 \\
\dpn{2002}{TC}{302} & 0.0239 & Estimated &  55.402 & 0.294 & 35.038 & 39.091 \\
\dpn{2010}{RF}{43}  & 0.0227 & Estimated &  49.417 & 0.246 & 30.643 & 37.253 \\
\dpn{2004}{GV}{9}   & 0.0218 & Estimated &  42.007 & 0.078 & 21.983 & 38.743 \\
\dpn{2002}{UX}{25}  & 0.0209 & Measured  &  42.736 & 0.143 & 19.433 & 36.643 \\
\dpn{2010}{KZ}{39}  & 0.0167 & Estimated &  45.267 & 0.056 & 26.089 & 42.724 \\
\dpn{2005}{UQ}{513} & 0.0121 & Estimated &  43.374 & 0.146 & 25.716 & 37.044 \\
\dpn{2012}{VP}{113} & 0.0118 & Estimated & 263.947 & 0.695 & 24.052 & 80.528 \\
\dpn{2014}{WK}{509} & 0.0105 & Estimated &  51.093 & 0.209 & 14.501 & 40.429 \\
\dpn{2005}{QU}{182} & 0.0088 & Estimated & 112.343 & 0.671 & 14.030 & 36.921 \\
\dpn{2010}{EK}{139} & 0.0054 & Estimated &  69.621 & 0.533 & 29.460 & 32.512 \\
\dpn{2002}{TX}{300} & 0.0018 & Measured  &  43.315 & 0.124 & 25.854 & 37.965
\enddata
\end{deluxetable*}

\section{Orbital Stability in the Outer Solar System}

\subsection{Introduction to Frequency Map Analysis}
\label{ss:freqan}

\begin{figure*}[ht!]
\plotone{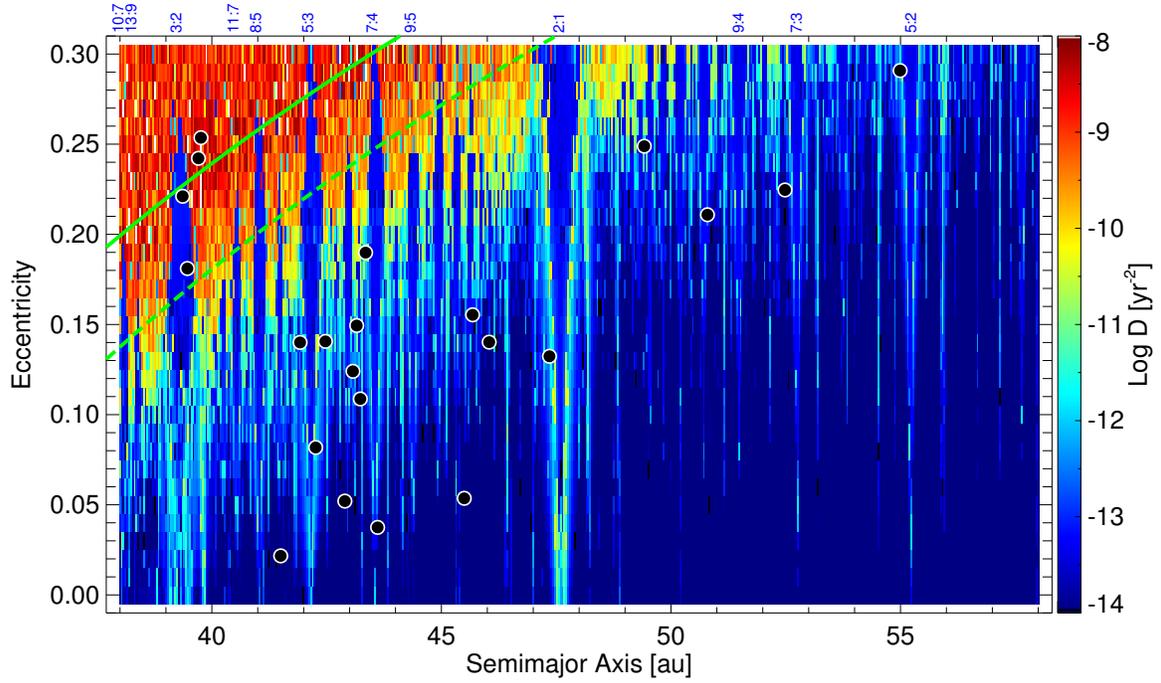}
\caption{Diffusion map for a wide region of the $a$ {\it vs} $e$ phase-space in the outer solar system. All the simulations were carried out with $i=0^\circ$. The color indicates the value of the logarithm of the diffusion parameter of the orbit in that location, which translates into the stability of orbits, i.e. bluer regions (lowest values of $D$) represent the more stable orbits, redder colors (highest values of $D$) represent the more unstable ones. The location and ratios of some MMRs with Neptune are labeled at the top of the figure. The solid green and dashed green lines indicate the perihelion distances at the location of Neptune and at $a_N+2\sqrt3R_{\it HN}$, respectively. Finally, the current location of 22 of the DPs in our set is marked by black circles.
\label{fig:innerdiff}}
\end{figure*}

\begin{figure*}[ht!]
\plotone{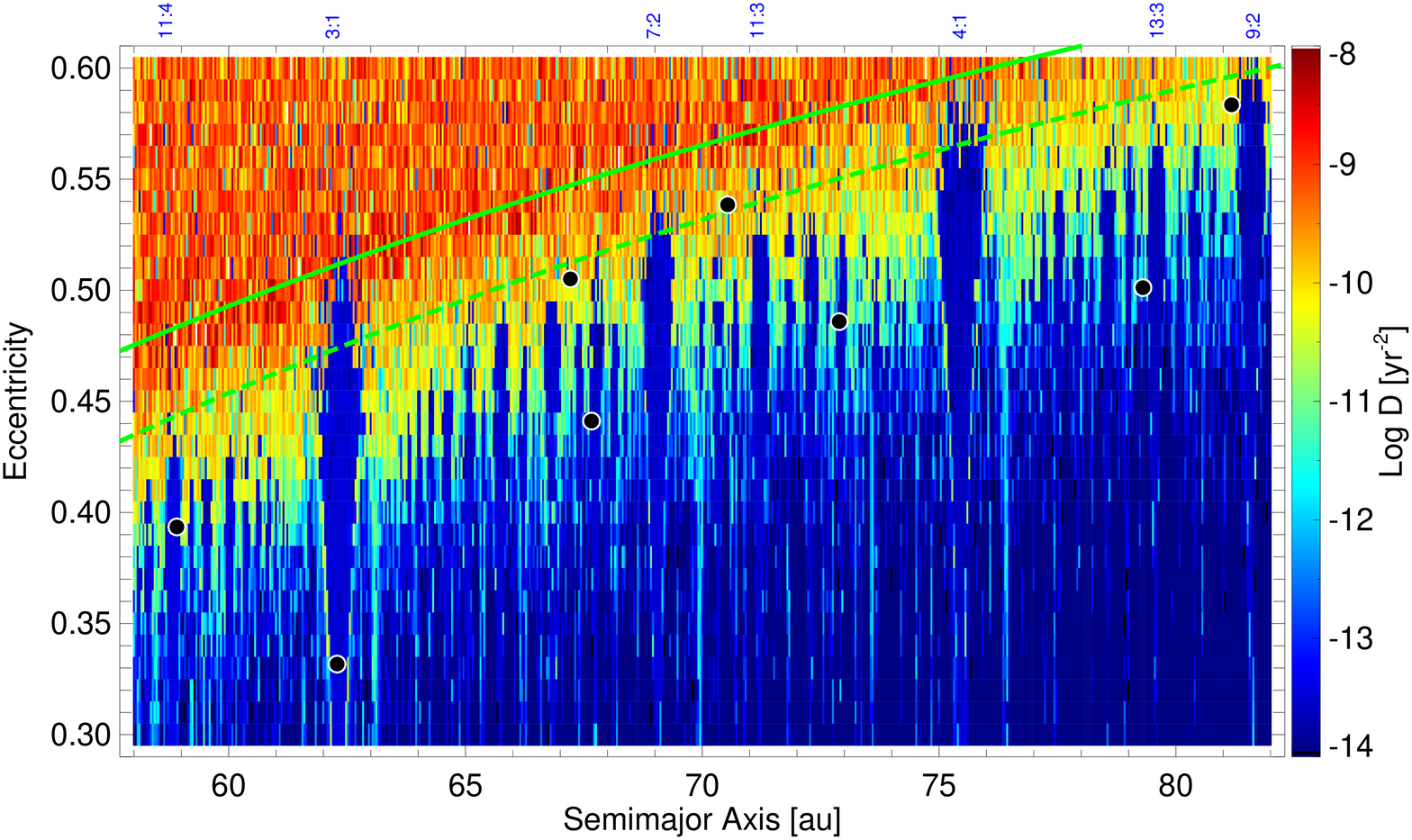}
\caption{Same as in Fig. \ref{fig:innerdiff} but for a farther out region of the $a-e$ plane. Here, the locations of another 8 DPs from our sample are marked by black circles. The remaining 4 DPs lie far outside the range of these Figures. 
\label{fig:outerdiff}}
\end{figure*}

We begin our analysis by characterizing the overall orbital stability of the outer solar system. To this aim, we performed a frequency map analysis \citep[FMA,][]{Laskar90,Laskar92} of wide regions of the $a$-$e$ phase space plane (semimajor axis vs eccentricity). A similar analysis has been performed earlier by \citet{Robutel01}. Here we reproduce their analysis but we increase the resolution in the sampling of the phase-space values corresponding to the outer solar system, focusing in the regions occupied by the majority of the DPs used in this work.

To produce the diffusion maps of Figs. \ref{fig:innerdiff}, \ref{fig:outerdiff}, \ref{fig:plutodiff}, and \ref{fig:fullmap}, we performed short-term numerical integrations (for the duration of $\sim1.31\,$Myr, or approximately 8000 orbital periods of Neptune) of thousands of test particles distributed in different homogeneous grids that cover the heliocentric $a$-$e$ phase-space plane.

The test particles are subject to the gravitational potential generated by the Sun and the four giant planets, where the mass of all the interior planets (including the Moon and Ceres) are added to that of the Sun. The initial conditions for all the massive bodies, on the Julian day 2458176.5 corresponding to February 27, 2018, were retrieved from the JPL's Horizons system\footnote{\url{https://ssd.jpl.nasa.gov/horizons.cgi}}. 

We create a diffusion map of the outer solar system by performing a frequency analysis of the evolution of each of our test particles; we focus on the quantity:
\begin{equation}
z(t)=a(t)\exp(i\lambda(t)),
\end{equation}
where $a$ and $\lambda$ are the semimajor axis and mean longitude of the particle, respectively, while the complex value of $z(t)$ represents a rough approximation of the position of the particle \citep[the details regarding the relevance of $z(t)$ can be found elsewhere, e.g.][]{Robutel01,Munoz17a}.

The algorithm used for the FMA was that of \cite{Sidly96}, which provides an approximate decomposition of $z(t)$ as follows:
\begin{equation}
z'(t)=\alpha_0\exp(i\nu_0t)+\sum_{k=1}^{N}\alpha_k\exp(i\nu_kt).
\end{equation}

For particles that lie in keplerian orbits $\lambda(t)=nt$ (where $n$ is the mean motion of the particle); also $N=0$, $a=|\alpha_0|$, and $n=\nu_0$ will remain constant. 

In general $N>0$, while $|\alpha_0|$ and $\nu_0$ need not remain constant; but, for moderately stable orbits, $|\alpha_0|$ and $\nu_0$ will remain nearly constant, and $a$ and $n$ will remain close to them (i.e. $\alpha_0\gg\alpha_k$); on the other hand, for unstable orbits these values will evolve quickly over time.

In particular $\nu_0$ (the mean frequency of the particle) will change for unstable orbits; a measure of the stability of the orbits can then be defined as the diffusion parameter:
\begin{equation} 
D=\frac{\left|\nu_{01}-\nu_{02}\right|}{T},
\end{equation}
where $\nu_{01}$ and $\nu_{02}$ are the main frequencies of a particle in each adjacent time interval of length $T$. Small values of $D$ indicate a stable trajectory, while larger values are indicative of unstable orbital evolution, resulting from strong variations of the main frequencies, which are characteristic of the orbital erratic nature. For a more detailed description of the FMA the interested reader is referred to the works by \citet{Laskar90,Laskar92,Laskar93}.

In order to compute the change of the main frequencies, we perform the frequency analysis for each particle in the intervals  $0<T<0.655$ Myr and $0.655<T<1.31$ Myr.

In the maps of Figs. \ref{fig:innerdiff}, \ref{fig:outerdiff}, and \ref{fig:fullmap} we color each ($a$,$e$) pair according to the logarithm of the diffusion parameter of its test particle such that redder colors indicate more unstable orbits, while bluer colors stand for the more stable orbits. Particles that are lost from the simulation before it finishes, mainly due to ejections or collisions with a planet, are colored white.

These diffusion maps let us identify regions where particles could survive in a long-term basis, while permitting us to easily identify the location and width of mean motion resonances with Neptune (MMRs). In the maps, the solid green lines indicate orbits with a perihelion $q=a_{\it Neptune}$ (i.e. it represents the crossing line of Neptune, with the implied risk of collision). The green dashed-lines indicate the distance from which the influence of Neptune starts to be dominant, as has been used in previous works \citep{Gladman90,Munoz19}; numerically the green dashed line can be expressed as: $q=a_N+2\sqrt{3}R_{\it HN}$, where $a_N$ and $R_{\it HN}$ are the semimajor axis and Hill radius of Neptune, respectively. 

\subsection{Stability and Resonances in the Plane}

Figs. \ref{fig:innerdiff} and \ref{fig:outerdiff} show the diffusion maps produced from the integration of $34\,162$ particles distributed in two homogeneous grids with the following conditions: in Fig. \ref{fig:innerdiff} the heliocentric $a$ is sampled from 38 to 58 au, with a step-size $\Delta a=0.04\,$au, while $e$ is sampled from 0 to 0.3 with a step-size $\Delta e=0.01$. For Fig. \ref{fig:outerdiff} $a$ is sampled from 58 to 82 au while $e$ covers from 0.3 to 0.6 in steps of the same size as in Fig. \ref{fig:innerdiff}.

In order to focus on the stability and shape of MMRs with Neptune, all of the remaining orbital elements: inclinations (from the plane of the ecliptic), $i$, arguments of pericenter, $\omega$, and mean anomalies, $M$, are set to zero. The longitudes of the ascending nodes, $\Omega$, are also nominally set to zero; while meaningless when $i=0^\circ$, they are a necessary input for the Mercury integrator.

The initial location of 30 out of the 34 DPs in our sample are indicated by black dots in both maps; the other 4 DPs all have $a>100$ au, and thus were left off of these figures.

\begin{figure}[ht]
\epsscale{1.178}
\plotone{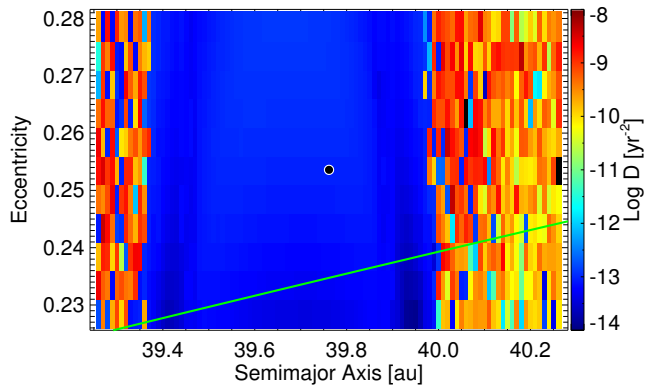}
\caption{Diffusion map around the orbit of Pluto (shown as a black circle). As in Fig. \ref{fig:innerdiff}, the color indicates the stability of the orbits. All test particles were run with the same inclination as Pluto, $i=17.12^\circ$. Note the sharp definition of the boundaries and the homogeneous stability of the 3:2 MMR. 
\label{fig:plutodiff}}
\end{figure}

We immediately notice that, for the majority of objects in our sample, a long-standing stability is expected based solely on their location on the phase-space. However these maps were made for an $i=0^{\circ}$; when studying orbits with $i\gtrsim 15^{\circ}$ the resonances move, slightly, outwards. We can see for example how two of the Plutinos (Pluto and Ixion) appear to be outside of the 3:2 MMR in Fig. \ref{fig:innerdiff}, even more, they appear to be in a highly unstable region. However, using a map with a more appropriate inclination it is easy to see that this is not the case; in Fig. \ref{fig:plutodiff} we present a small diffusion map using $i=17.12^{\circ}$, the heliocentric inclination of Pluto; this map clearly shows that Pluto is firmly inside the 3:2 resonance. The shifting effect of the location of MMRs, due to the change in $i$, will occur for all resonances; for example, we can see the opposite case for \dpn{2015}{KH}{162}: in Fig. \ref{fig:outerdiff} it appears to be clearly located inside the 3:1 MMR, however, we were unable to identify a stable libration of any of the possible second-order resonant arguments for it; therefore, despite its suggestive location on the map, due to its high inclination, \dpn{2015}{KH}{162} is found to be clearly non-resonant.

Based on the diffusion maps, it is possible to understand the low stability of some of the DPs in our simulations, for example, \dpn{2007}{OR}{10} and \dpn{2015}{RR}{245}, are very clearly located in the broad unstable region (close to the dashed line: $a_{N}+2\sqrt{3}R_{\it HN}$), while \dpn{2002}{MS}{4} and \dpn{2002}{UX}{25} are located in the unstable regions near both sides of the 5:3 MMR. On the other hand \dpn{2005}{RN}{43}, Quaoar, Varuna, and \dpn{2010}{KZ}{39} represent examples of objects with low $e$ located in very stable regions.

\subsection{Stability and Diffusion in the Kuiper Belt with Random Inclinations}

\begin{figure*}[ht!]
\plotone{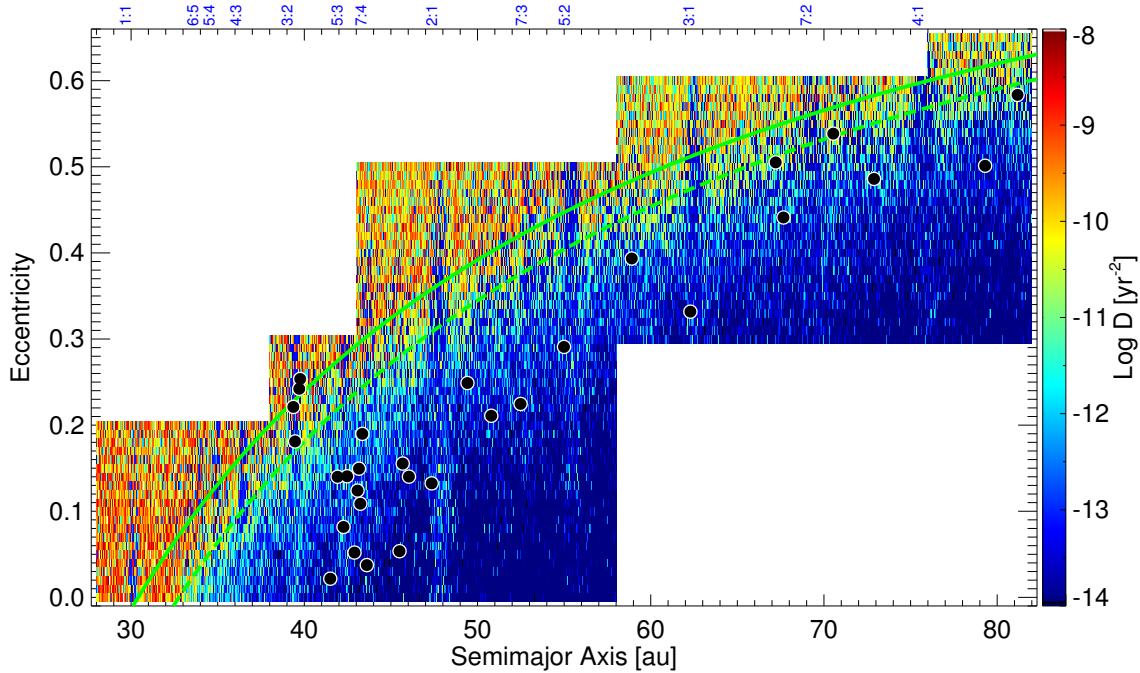}
\caption{Similar to Figure \ref{fig:innerdiff}. Diffusion map for the widest region of the $a-e$ phase-space plane considered in this work. For this plot, the inclinations of test particles were chosen randomly between 0$^\circ$ and 50$^\circ$. \label{fig:fullmap}}
\end{figure*}

\begin{figure}[ht!]
\epsscale{1.178}
\plotone{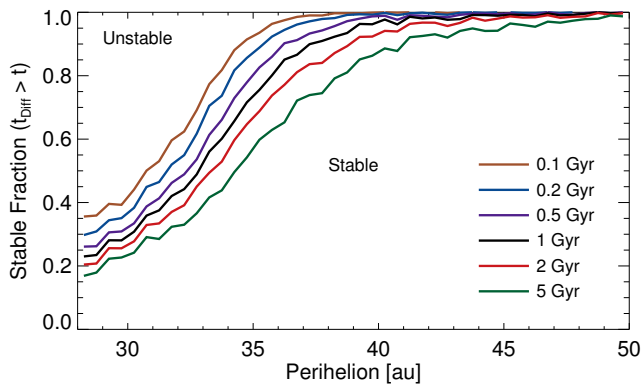}
\caption{Fractions of stable particles for different time-scales as a function of their perihelion. Each line shows the expected fraction of the population of the Kuiper belt that would be stable from 0.1 to 5 Gyr, as a function of the perihelion distance. For example, beyond a perihelion of 41 au, more than 90\% of the trans-Neptunian populations are expected to remain stable for the remaining lifetime of the solar system. 
\label{fig:fracuns}}
\end{figure}

In order to better understand the effect that a non-zero inclination would have on the general aspect and on the stability picture drawn from the diffusion maps, we ran a set of simulations including $48\,235$ total test particles covering different patches of the $a$-$e$ phase space. In this case, the inclination of each particle was assigned at random, with values between 0$^\circ$ and 50$^\circ$, in order to represent the broad distribution of inclinations in our DP sample (where the minimum and maximum $i$ values are 7.552$^\circ$ and 43.993$^\circ$, respectively). Besides, the choosing of such values allow us to better identify the resonances for objects that orbit far away from the plane.

Since each of the 34 DPs also has its own value for the other angular parameters, $\omega$, $\Omega$, and $M$, we decided to assign random values between 0$^\circ$ and 360$^\circ$ for the corresponding initial conditions of each test particle in the map. The exact values used for these angles are probably not very important; but, where there is any difference, this approach has the advantage of showing us, locally, the range of stability available to objects with such $a$-$e$ values. We will study a more representative determination for each of the objects of our sample in section \ref{ssec:SpecSta}.

In Fig. \ref{fig:fullmap} we show the resulting diffusion map for random inclinations and angular elements, as described above, for a larger region of the phase-space than those of Figs. \ref{fig:innerdiff} and \ref{fig:outerdiff} (though both these regions are covered in the simulations of this section). As in the previous maps, the solid green line delimits the collision region with Neptune, while the dashed green line delimits the region where Neptune's gravity is dominant. The location of 30 out of 34 DPs is again marked by black dots.

It is interesting to note how the general form of the MMRs changes with respect to the zero inclination case. With random angles and inclinations, the definition of the resonances gets blurred; however, the stronger MMRs, whose ratios are labeled at the top of the figure, are still easily identifiable. It is evident that the locations of the resonances are perturbed by the value of the inclination, thus we can see how both Pluto and Ixion are indeed located within the limits of the 3:2 MMR.

This diffusion map constitutes a better representation of the stability of the Kuiper belt as a whole, not only for the DPs but for objects of all sizes in the various dynamical families present in the region covered by the map (namely the Classical belt, the Scattered disk and some Resonant populations). From the map, it is possible to estimate, statistically, the diffusion time of Kuiper belt objects as a function of their perihelion. 

In Figure \ref{fig:fracuns} we present the fraction of stable orbits as a function of their perihelion, for diffusion times varying between 0.1 to 5 Gyr. The diffusion time, $T_\mathrm{Diff}$, is obtained from the diffusion parameter as $T_\mathrm{Diff}=1/(DP)$, where P is the period of the orbit \citep[see for example][for other recent applications of the diffusion time on different systems]{Gaslac20,Roberts21}. To determine the stable fraction, we counted the number of particles, within our sample, for each perihelion band between 28 and 50 au, whose orbits have diffusion times above 6 different time limits: 0.1, 0.2, 0.5, 1, 2, and 5 Gyr. These limits are significant in the sense that any object with a diffusion time below 0.1 Gyr is expected to be lost very quickly, and thus those orbits are not expected to have a meaningful contribution to the present day population; on the other hand a diffusion time above 5 Gyr represent longer than the lifetime of the solar system.

We note that our sample is incomplete; besides having a hard limit at a semimajor axis of 82 au, there are additionally a few gaps below 30 au and above 40.6 au. Nonetheless, the sample is numerous enough as to provide a good representation of the expected stable fraction of Kuiper belt objects as a function of perihelia. For example, we expect that, from an initial population with perihelia of 38 au: almost all objects would be stable for more than 100 Myr, 10\% of this initial population would be unstable in time scales of 1 Gyr, and nearly 30\% of the population would be lost over the age of the solar system.

\subsection{Specific Stability of the Objects of our DP Sample \label{ssec:SpecSta}}

\begin{deluxetable*}{lDDDc}
\tablecaption{Diffusion values for the 34 DPs}
\label{tab:DpsDiff}
\tablewidth{0pt}
\tablehead{
\colhead{} & \multicolumn{2}{c}{Log($D$)} & \multicolumn{2}{c}{$P$} & \multicolumn{2}{c}{$T_{\it Diff}$} & \colhead{Predicted} \vspace{-5pt}
\\ 
\colhead{Object} & \multicolumn{2}{c}{[yr$^{-2}$]} & \multicolumn{2}{c}{[yr]} & \multicolumn{2}{c}{[Gyr]} & \colhead{Stability} 
}
\decimals
\startdata
      Eris & -13.5641 &   556.85 &    65.822 & Very Stable \\
     Pluto & -13.1178 &   248.80 &    52.711 & Very Stable \\
    Haumea & -11.3346 &   284.99 &     0.758 & Unstable \\
\dpn{2007}{OR}{10} & -11.5276 &   546.80 &     0.616 & Unstable \\
  Makemake & -13.5108 &   309.76 &   104.658 & Very Stable \\
    Quaoar & -13.6914 &   285.61 &   172.049 & Very Stable \\
\dpn{2002}{MS}{4} & -11.7144 &   269.35 &     1.924 & Unstable \\
     Sedna & -15.4648 & 11674.73 &   249.791 & Very Stable \\
     Orcus & -14.1432 &   248.08 &   560.590 & Very Stable \\
\dpn{2014}{EZ}{51} & -13.6885 &   377.34 &   129.356 & Very Stable \\
\dpn{2010}{JO}{179} & -13.5882 &   702.34 &    55.161 & Very Stable \\
\dpn{2002}{AW}{197} & -12.8564 &   327.59 &    21.933 & Very Stable \\
\dpn{2015}{KH}{162} & -14.2402 &   489.98 &   354.823 & Very Stable \\
     Varda & -12.8356 &   309.75 &    22.108 & Very Stable \\
\dpn{2007}{UK}{126} & -11.6136 &   631.89 &     0.650 & Unstable \\
\dpn{2013}{FY}{27} & -12.5621 &   453.66 &     8.042 & Stable \\
\dpn{2003}{AZ}{84} & -12.2695 &   249.58 &     7.453 & Stable \\
\dpn{2015}{RR}{245} & -13.7521 &   736.19 &    76.747 & Very Stable \\
\dpn{2003}{OP}{32} & -13.6263 &   282.22 &   149.884 & Very Stable \\
\dpn{2014}{UZ}{224} & -13.8002 &  1146.20 &    55.069 & Very Stable \\
     Ixion & -12.9618 &   248.25 &    36.887 & Very Stable \\
    Varuna & -13.7116 &   283.30 &   181.708 & Very Stable \\
\dpn{2005}{RN}{43} & -13.8845 &   265.81 &   288.369 & Very Stable \\
\dpn{2002}{TC}{302} & -12.3153 &   410.98 &     5.029 & Stable \\
\dpn{2010}{RF}{43} & -12.5647 &   344.67 &    10.648 & Stable \\
\dpn{2004}{GV}{9} & -12.9266 &   274.10 &    30.813 & Very Stable \\
\dpn{2002}{UX}{25} & -13.1511 &   277.98 &    50.942 & Very Stable \\
\dpn{2010}{KZ}{39} & -13.5790 &   304.64 &   124.518 & Very Stable \\
\dpn{2005}{UQ}{513} & -13.2839 &   283.94 &    67.717 & Very Stable \\
\dpn{2012}{VP}{113} & -16.9210 &  4166.75 & 20006.962 & Very Stable \\
\dpn{2014}{WK}{509} & -13.3754 &   364.86 &    65.059 & Very Stable \\
\dpn{2005}{QU}{182} & -13.8987 &  1160.03 &    68.272 & Very Stable \\
\dpn{2010}{EK}{139} & -12.0525 &   584.02 &     1.932 & Unstable \\
\dpn{2002}{TX}{300} & -13.1605 &   283.03 &    51.126 & Very Stable 
\enddata
\end{deluxetable*}

In order to obtain a better estimate, regarding the specific stability of each object in our sample of large TNOs (i.e. beyond the global stability suggested by the general configuration of the diffusion map of Fig. \ref{fig:fullmap}), we performed additional stability simulations of 100 mass-less clones of each DP. The clones were generated by randomly assigning orbital elements within the intervals $a_{\it DP}\pm0.04$ au and $e_{\it DP}\pm0.01$, while the angular elements, $i_{\it DP}$, $\omega_{\it DP}$, $\Omega_{\it DP}$, and $M_{\it DP}$, were randomly generated within intervals of $\pm1^\circ$ around the initial values of each DP's orbit.

We calculated the value of the diffusion parameter for each DP, as the average of the parameters of its 100 clones. The results are shown in Table \ref{tab:DpsDiff}. In the last column of Table \ref{tab:DpsDiff}, we give a prediction on the stability of each object based on the value of its diffusion time. 


Our definition of $T_\mathrm{Diff}$ represents an estimate of the time scale for a change of unity in the period of an orbit; from Kepler's Third Law it is clear that changes in period (or mean motion) originate from changes in the semimajor axis of the orbit. In this work we assume that a drastic change in the orbital motion would be a quarter of this, and we will set our criteria by comparing our stability time scales with $T_\mathrm{Diff}/4$.


Based on the previous argument, we call ``Very Stable'' those orbits with $T_\mathrm{Diff} > 20$ Gyr, this is, those for which a significant change in their mean motion would not occur over the remaining life of the solar system. We call ``Stable'' those orbits for which $4<T_\mathrm{Diff}<20$ Gyr, this is, an appreciable change in their orbits can occur over the age of the solar system, but requiring more than 1 Gyr. Finally, ``Unstable'' are those orbits with $T_\mathrm{Diff}<4$ Gyr, i.e. orbits which will experience drastic changes in less than 1 Gyr of evolution.

The 1 Gyr limit may seem arbitrary, however, many dynamical works which consider a time scale for the study of continuous and stationary phenomena in the solar system, a so called steady state, use this same scale of 1 Gyr \citep[e.g.][]{Levison97,Tiscareno09,Nesvorny17,Yu18,Nesvorny19}. Indeed, the present study arises from 1 Gyr integrations, which were performed for the characterization of the contribution of DPs to the delivery of short-period comets, from the Kuiper belt to the inner solar system \citep{Munoz19}. We will present the results for the DPs of those simulations, as well as several complementary 1 Gyr integrations, in the next sections.

From Table \ref{tab:DpsDiff} we can see that \dpn{2007}{OR}{10} is the DP with fastest diffusion rate, with important changes in a time scale of 150 Myrs. Also, Haumea, \dpn{2002}{MS}{4}, \dpn{2007}{UK}{126}, and \dpn{2010}{EK}{139} have significant changes in less than 1 Gyr.

On the other extreme we find that the diffusion time scale for \dpn{2012}{VP}{113} is greater than 5000 Gyr (5 Tyr!). In fact 21 DPs (out of 34) have stabilities beyond 5Gyr (beyond the remaining lifetime of the solar system).

\section{Long term dynamical simulations}

While illuminating and helpful in providing a general picture of the stability of the outer solar system, the short-term simulations of the previous sections only provide us with a prediction on the stability of the objects located in those regions. It is evident that this is not enough to observe the evolution of the orbits and their possible departure from their current regions of regular or quasi-regular evolution. On top of that, the previous simulations lacked the simultaneous presence of all 34 DPs, thus ignoring the effect that they can have on each other.

We know from previous works that, while cometary nuclei in the Kuiper belt are generally stable, there is a constant supply of new objects towards the inner solar system, sustained by instabilities in the trans-Neptunian reservoirs \citep[see][for a recent review on comets and their reservoirs]{Dones15}.

From the previous arguments, it becomes clear that the proper study of the stability of the objects in the outer solar system requires two more things: longer integrations and a more complete dynamical model of the solar system. 

To this aim, we will take advantage of 1~Gyr long simulations we ran to study the fate of cometary nuclei in the presence of the 34 largest trans-Neptunian DPs \citep{Munoz19}. A by-product of those simulations is a large amount of data on the fate of each individual DP, which will let us study their dynamical evolution on a statistical basis.

\subsection{Dynamical Model of Our TNO Sample}
\label{ssec:dynmodel}

Our dynamical model is composed, in all cases, by a central star with a mass equal to $(1+\epsilon)$M$_\odot$, where $\epsilon$ is a small quantity that represents the mass of all the terrestrial planets plus the Moon and Ceres; also, we consider the gravitational influence of the 4 giant planets and the 34 massive DPs.

We ran a total of 200, 1~Gyr long simulations, using the hybrid symplectic integrator from the Mercury package \citep{Chambers99}. The initial conditions were the same as in section \ref{ss:freqan}.

Most of the simulations we are using (177 out of 200) were designed for the study of the evolution of cometary nuclei and each one included a different set of some hundreds of test particles; the results for the evolution of test particles on many of these simulations were reported in \citet{Munoz19}.

The initial conditions of the massive objects in all 177 simulations were identical, but each simulation differs from the others by the specific set of test particles which were included; while the nominal time step for all integrations was 400 days, the presence of close interactions forced the integrator to modify the time steps when required, this implies that, each specific set of test particles resulted in a different specific set of time steps for each simulation, which in turn resulted in different evolutions for the massive objects in each case. The differences in evolution are barely noticeable for the giant planets, but sometimes resulted in wildly different positions for the DPs; since the error tolerances for the integrations were stringent (with a tolerance accuracy parameter of $10^{-10}$), these differences represent the evolution of very close orbits in a chaotic region of the phase-space, and indeed all these results are consistent with the very same initial conditions within very small error bars.

To the previous sample we added 23 extra simulations, to end up with an even 200. Each of the new simulations had the same initial conditions for the massive objects, and included a token test particle at 100 au; instead of depending on interactions on test particles, which slow down the integrations, each new simulation has a slightly different initial time-step (between 389-412 days), which again resulted in different evolutions for each of our DPs.

\subsection{Validity and Meaning of the Different Results}

The statistical studies we present here are possible due to the different results obtained in different simulations, despite the identical initial conditions used for the massive objects. Those differences could na\"ively put into question the validity of our results, or be attributed simply to the finite precision of the integrator; however, our usage of a highly reliable and widely accepted integrator \citep[the Mercury package of][with over 1000 citations]{Chambers99}, could easily quell at least the later argument.

On the other hand, it is understood that numerical errors are unavoidable in integrations and even more, while studying a very unstable system, any small imprecision may lead to important divergences over time, i.e. dynamical chaos \citep[e.g.][]{Murray99}. This doesn't invalidate the meaning of such integrations; in fact, the numerical errors in the short-term are smaller that the uncertainties in the observational data for the initial conditions of the objects. On the longer term, however, one could argue, that only the very stable orbits will exist in the real system, since any unstable orbit would be expected to have been expelled from the system long ago; leading to the false conclusion that the divergences found in our numerical experiments cannot have an observational equivalent nowadays. On the face of this argument, we must remember that the solar system has thousands of objects that we are not including in our simulations, which can chaotically interact at different levels, and a small interaction with another close or even not so close component may lead to a divergence more significant that those produced by numerical errors.

To these we must add the presence of extra solar perturbations which are not easily predictable neither accounted for, such as: moderately close stellar flybys or interstellar visitors ('Oumuamua, 2I/Borisov); any of which also can destroy the stable orbits expected in a perfectly steady-state system, or the ones produced by a perfect hypothetical integrator. 

In a sense, the instability of orbits in the trans-Neptunian region is hardly unexpected, since it has been long known that some apparently stable orbits, over time spans of Gyr, can suddenly and quickly become chaotic on time spans of Myr \citep{Torbett89,Holman93,Duncan95}; this phenomenon supplies comets to the inner solar system. At first glance the instability of larger objects, such as a DP, may seem surprising, however, there is nothing that prevents, in principle, such massive objects from suffering the same fate as cometary nuclei, at least in a statistical sense. To put it simply, the constant presence of new comets shows that the solar system, and in particular the trans-Neptunian region, is not in a perfectly smooth and stable steady-state.

Of course, we do not expect all DPs to be unstable (see Table \ref{tab:DpsDiff}, as well as Section \ref{sec:res}); some of the most stable orbits (such as Pluto's) will not be destabilized by inaccuracies resulting form numerical errors. Still, for objects surrounded by a not-so stable phase space, the presence of slight (but constant) perturbations can disrupt their stability. 

\section{Results and Discussion}
\label{sec:res}

\subsection{Classification and global results}
\label{subsec:class}

\begin{figure*}
    \includegraphics[width=\linewidth]{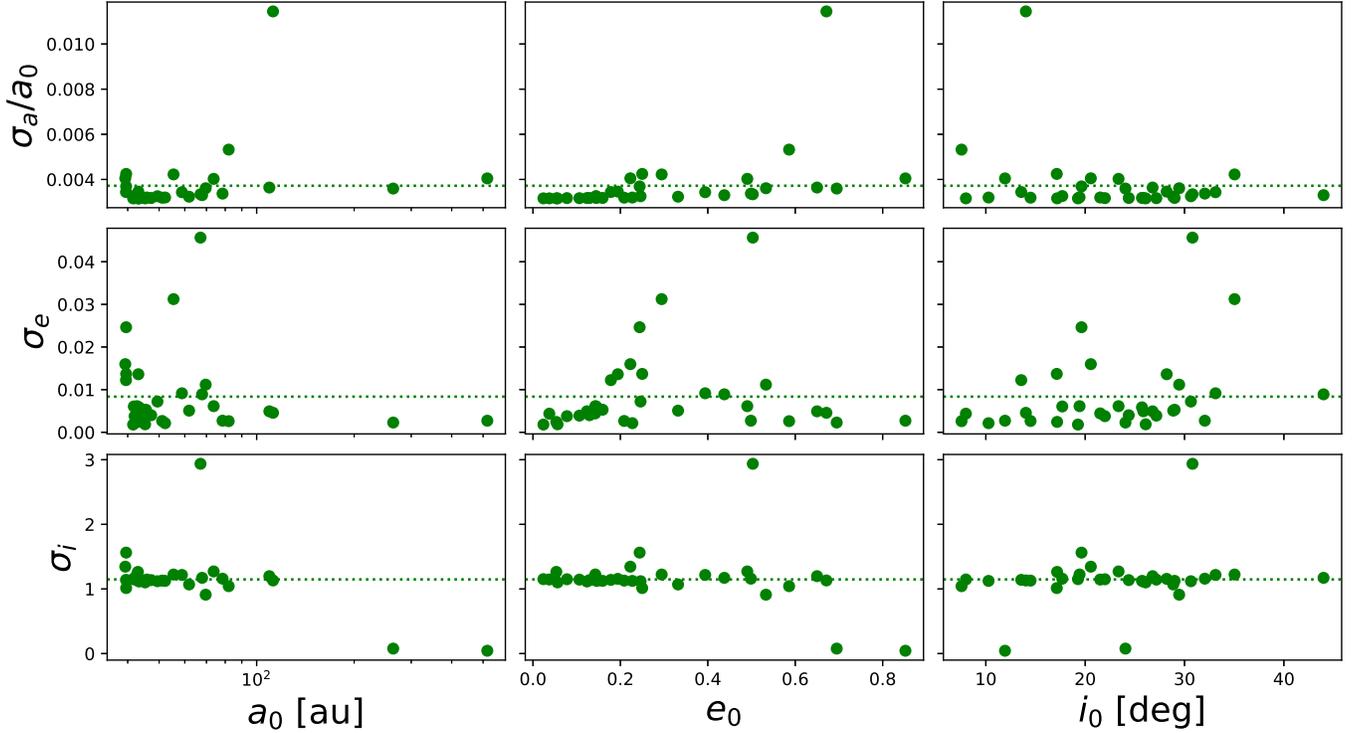}
\caption{Standard deviations of the 200 realizations for each DP after 25 Myr of evolution. Green circles are the $\sigma$'s for each of the main orbital parameters against the initial conditions for each DP. A dotted green line marks the average of all the points in each panel, which correspond to one fifth of our reference value, used to differentiate between regular and irregular orbits. \label{fig:sigmas}}
\end{figure*}

\begin{deluxetable*}{lDccccc}
\tablecaption{1 Gyr stability for the 34 DPs}
\label{tab:sigmas}
\tablewidth{0pt}
\tablehead{
\colhead{} & \multicolumn{2}{c}{M} &\multicolumn{4}{c}{Regular Orbits} & \colhead{} \vspace{-5pt}\\
\colhead{Object} & \multicolumn{2}{c}{[$10^{-3}$ \mearth]} & \colhead{a} & \colhead{e} & \colhead{i} & \colhead{all 3} & \colhead{Classification}
}
\decimals
\startdata
Eris                & 2.7956 & 197 & 191 & 197 & 191 & Stable   \\
Pluto               & 2.4467 & 200 & 200 & 200 & 200 & Resonant (3:2) \\
Haumea              & 0.6706 &  46 &   4 &  53 &   4 & Unstable \\
\dpn{2007}{OR}{10}  & 0.6110 &  58 & 128 & 145 &   0 & Unstable \\
Makemake            & 0.5531 & 199 & 187 & 200 & 187 & Stable   \\
Quaoar              & 0.2343 & 200 & 200 & 200 & 200 & Stable   \\
\dpn{2002}{MS}{4}   & 0.1369 & 129 &  95 & 133 &  95 & Unstable \\
Sedna               & 0.1254 & 200 & 200 & 200 & 200 & Stable   \\
Orcus               & 0.1073 & 161 &   1 & 153 &   1 & Unstable \\
\dpn{2014}{EZ}{51}  & 0.1012 & 200 & 197 & 200 & 197 & Stable   \\
\dpn{2010}{JO}{179} & 0.0635 & 200 & 188 & 200 & 188 & Resonant (21:5) \\
\dpn{2002}{AW}{197} & 0.0606 & 200 & 199 & 200 & 199 & Stable   \\
\dpn{2015}{KH}{162} & 0.0594 & 200 & 200 & 200 & 200 & Stable   \\
Varda               & 0.0446 & 200 & 200 & 200 & 200 & Stable   \\
\dpn{2007}{UK}{126} & 0.0415 &  66 &  95 & 102 &  59 & Unstable \\
\dpn{2013}{FY}{27}  & 0.0391 & 122 &  82 & 130 &  82 & Unstable \\
\dpn{2003}{AZ}{84}  & 0.0349 & 197 & 188 & 196 & 188 & Resonant (3:2) \\
\dpn{2015}{RR}{245} & 0.0331 &  14 &  16 &  67 &  14 & Unstable \\
\dpn{2003}{OP}{32}  & 0.0288 & 200 & 200 & 200 & 200 & Stable   \\
\dpn{2014}{UZ}{224} & 0.0266 &  97 & 180 & 163 &  88 & Unstable \\
Ixion               & 0.0263 & 200 &   0 & 196 &   0 & Resonant (3:2) \\
Varuna	            & 0.0259 & 200 & 200 & 200 & 200 & Stable   \\
\dpn{2005}{RN}{43}  & 0.0255 & 200 & 200 & 200 & 200 & Stable   \\
\dpn{2002}{TC}{302} & 0.0239 & 188 &   0 & 151 &   0 & Resonant (5:2)\\
\dpn{2010}{RF}{43}  & 0.0227 & 186 & 170 & 190 & 170 & Unstable \\
\dpn{2004}{GV}{9}   & 0.0218 & 200 & 168 & 200 & 168 & Stable   \\
\dpn{2002}{UX}{25}  & 0.0209 & 199 & 173 & 199 & 173 & Unstable \\
\dpn{2010}{KZ}{39}  & 0.0167 & 200 & 200 & 200 & 200 & Stable   \\
\dpn{2005}{UQ}{513} & 0.0121 & 200 & 199 & 200 & 199 & Stable   \\
\dpn{2012}{VP}{113} & 0.0118 & 200 & 200 & 200 & 200 & Stable   \\
\dpn{2014}{WK}{509} & 0.0105 & 200 & 200 & 200 & 200 & Stable   \\
\dpn{2005}{QU}{182} & 0.0088 &   0 &  98 & 137 &   0 & Unstable \\
\dpn{2010}{EK}{139} & 0.0054 & 198 &   2 &   2 &   2 & Resonant (7:2) \\
\dpn{2002}{TX}{300} & 0.0018 & 200 & 200 & 200 & 200 & Stable   \\
\enddata
\end{deluxetable*}

We now present the statistical results from our long-term simulations; this is, the results from 200 different realizations for each of the 34 DPs (for clarity, in the rest of this work, we will call "orbit" each of the 200 realizations for the evolution of each DP; i.e. we will be studying 6800 different orbits). As we did in section \ref{ssec:SpecSta}, we are interested in providing a general classification for each object (stable or unstable), as well as their possible evolutions, since, as we will show, some of the DPs are likely to leave the trans-Neptunian region during the remaining lifetime of the Sun. 

An ideally stable object would present 200 regular orbits. This is, each orbit would have only small variations on its main orbital parameters during our entire simulation; this also implies that the orbits on all of the 200 simulations would be equivalent. 

In order to define what a regular (or irregular) orbit is, we need to determine the limits in the deviations of the orbital parameters that we will tolerate and still call an orbit ``regular''. We first calculate the standard deviations of the orbital parameters $a$, $e$, and $i$ in the first 25~Myr of each integration; this is, for each object we calculate the $\sigma_a$, $\sigma_e$, and $\sigma_i$ using the data of the first 25~Myr for all 200 of its orbits. In Fig. \ref{fig:sigmas} we plot the values of $\sigma_a/a_0$, $\sigma_e$, and $\sigma_i$ versus the initial orbital elements of each DP, this is, against $a_0$, $e_0$, and $i_0$. From Fig. \ref{fig:sigmas} we can see that most of the DPs have similar values of their corresponding dispersion. Using the values of Fig. \ref{fig:sigmas} we now define a characteristic deviation for each orbital element simply as the average of the above standard deviations, thus we have $\left<\sigma_a/a_0\right>=0.00372$, $\left<\sigma_e\right>=0.00836$, and $\left<\sigma_i\right>=1.148^\circ$. With these characteristic values, we are now able to determine, for each DP, how many of the orbits are regular in the corresponding parameter; we do this by defining an individual orbit as regular if its orbital parameters ($a$, $e$, and $i$) remain within 5 characteristic deviations from their original value for the entire simulation, i.e. they remain within the intervals $a_0\left(1\pm5\left<\sigma_a/a_0\right>\right)$, $e_0\pm5\left<\sigma_e\right>$, and $i_0\pm5\left<\sigma_i\right>$ for 1 Gyr. Note that we do not use any restrictions on the evolution of $\omega$, $\Omega$, or $M$.

In Table \ref{tab:sigmas} we show the number of regular orbits in each orbital parameter (out of 200), according to the previous procedure. We also include the number of orbits which were regular in all 3 parameters simultaneously. We will consider a DP to be stable if it fulfills the following two conditions: if at least 80\% of its orbits are stable, and if none of its orbits ended up being lost in the simulation. We consider 3 kinds of losses: objects whose $a$ increases beyond $10\,000$~au (escape from the solar system), objects whose distances to the sun decrease under 1 au (big comet), or collisions with a planet. 

Of the 15 objects classified as unstable in Table \ref{tab:sigmas}, 13 were classified as such for having between 0 and 95 stable orbits (clearly less than 80\%); the other two unstable objects, \dpn{2010}{RF}{43} and \dpn{2002}{UX}{25}, were classified as such since, despite having 170 and 173 stable orbits, some of their orbits were lost (5 and 1, respectively).

An additional result of the analysis of unstable and semi-stable orbits is the presence of resonant orbits; we have added an additional category to allow for the specifics of the behavior of such orbits. There are 6 objects in Table \ref{tab:sigmas} that have been classified as resonant; 3 of those (Pluto, \dpn{2010}{JO}{179}, and \dpn{2003}{AZ}{84}) could easily be considered as stable, while the other 3 (Ixion, \dpn{2002}{TC}{302}, and \dpn{2010}{EK}{139}) would be considered unstable according to our previous selection criteria. It is worth to note that, although the last 3 objects could be classified as unstable, they are not really fully unstable, as they remain locked in their resonance 94\% - 100\% of the time (as shown by the number of orbits regular in $a$). The reduced number of regular orbits for such objects (between 0 and 2) is actually due to their resonant nature, which leads to large $e$ variations (and sometimes of the inclination) above the 5-$\sigma$ level defined before. This new category highlights their resonant behaviour which dominates for most of their orbits and is the most defining characteristic of their evolution; it is for this reason that we have also included the three more-stable resonant objects in this category.


\begin{figure}[ht!]
\epsscale{1.178}
\plotone{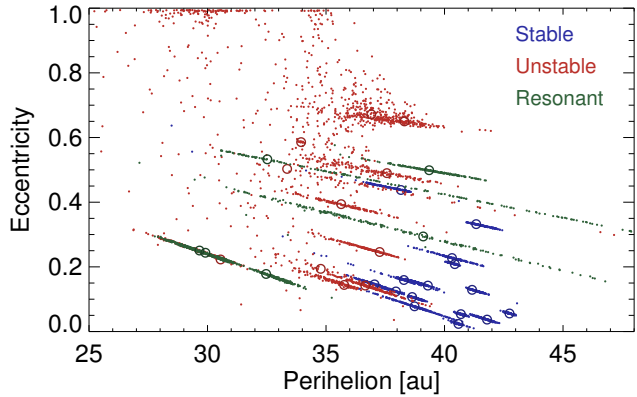}
\caption{Scattering of final conditions in the eccentricity-perihelion plane. The colored dots (blue for stables, red for unstables, and green for resonants) mark the final values on each of the 200 realizations for 32 DPs (Sedna and \dpn{2012}{VP}{113} perihelia lie far beyond the range of the figure). Colored open circles indicate the initial condition of each DP in this plane. \label{fig:qvse}}
\end{figure}

In figure \ref{fig:qvse} we present the initial and final perihelion vs eccentricity for all 200 orbits for 32 DPs (Sedna and \dpn{2012}{VP}{113}, have perihelion of 76.2 and 80.5 au, outside the range of this figure). We can see that for some DPs (the stable ones; blue) the final values for each orbit (dots) lie all very close to one another, and also lie very close to the initial condition (circle); for the resonant DPs (green) we see that the initial and final values lie in long lines corresponding to constant $a$, showing the evolution in $e$ common in resonant orbits; for both stable and resonant orbits there are frequently a few stragglers far from the initial values. On the other hand there are DPs (unstable DPs; red) where the final points change both in $e$ and $a$, sometimes spreading over large areas of this plane. This highlights the differences of the 3 classifications defined for Table \ref{tab:sigmas}.

The last column in Table \ref{tab:sigmas} gives us an idea of the quality of the determinations presented in Table \ref{tab:DpsDiff}. Of the 28 non-resonant objects, 21 remain in the same category (considering ``Very Stable" objects to be in the ``Stable" category). Of the other 7 objects: \dpn{2010}{RF}{43} changes, but it is near the limit of both classifications, while the other 6 change category completely (Orcus, \dpn{2013}{FY}{27}, \dpn{2015}{RR}{245}, \dpn{2014}{UZ}{224}, \dpn{2002}{UX}{25}, \dpn{2005}{QU}{182}); in all 7 cases the new determination is unstable while the old one was either ``Stable" (2 cases) or ``Very Stable" (5 cases). This difference occurs mostly because of the limitations of the FMA, which extrapolates the amount that the orbit will change based on a short-term integration in which the orbit does not deviate substantially from its initial conditions, i.e. while the perturbations are small their behaviour can be modeled by the FMA, but when they grow large enough to reach the direct influence of Neptune (within 3 hill radii of Neptune) the stability decreases dramatically and the short term estimates are no longer predictive.

Regarding the resonant DPs: we find that Pluto, \dpn{2010}{JO}{179}, and \dpn{2003}{AZ}{84} were stable in the FMA and remain stable in the long-term integration. On the other hand Ixion and \dpn{2002}{TC}{302} have regular evolution for $a$ and $i$ but their $e$ changes drastically (as is expected for many resonant objects). Finally \dpn{2010}{EK}{139} was unstable in the FMA and shows erratic behavior, in both $e$ and $i$, in the long-term integrations while remaining in the resonance (with constant $a$).


The 4 Plutinos represent an interesting subset of DPs, showing a wide variety of behaviors. While Pluto is completely (100\%) stable, \dpn{2003}{AZ}{84} would be considered stable (94\%), and Ixion would be considered unstable (0\% stable); yet, out of the 600 orbits corresponding to these 3 DPs only 3 orbits leave the resonance. Orcus, on the other hand, is truly unstable; not only 39 orbits leave the resonance (which, in itself negates the argument that it is locked in the 3:2 resonance); but, out of those 39 orbits, 26 leave the simulation entirely (18 leaving the solar system and 8 falling to the inner solar system). 

A final note should be made about the definition of stability: one could measure stability by the consistency in $e$ and $i$, or by the consistency of the orbital period (or $a$), or by the ability to remain in the Kuiper belt; the least stable object would be different in each case. Ixion is the DP with the smallest amount of regular orbits (0\%) while remaining in resonance, \dpn{2005}{QU}{182} is the DP that is less likely to remain close to its original $a$ (0\%) while Haumea is the most unstable DP considering the number of orbits that remain in the solar system (37.5\%).

\subsection{Different outcomes for irregular orbits}
\label{subsec:outcomes}

\begin{deluxetable}{lccccc}
\tablecaption{DPs with ejections}
\label{tab:ejects}
\tablewidth{0pt}
\tablehead{
\colhead{Object} &  \colhead{\thead{Total\\ Ejections}} & \colhead{\thead{Outer\\ Space}} & \colhead{\thead{Inner\\ System}} & \colhead{Collisions}
}
\decimals
\startdata
Haumea              & 125 & 98 & 25 & 2(Jup) \\
\dpn{2007}{OR}{10}  &  80 & 70 &  9 & 1(Nep) \\
\dpn{2002}{MS}{4}   &  51 & 41 & 10 & 0 \\
\dpn{2015}{RR}{245} &  40 & 33 &  6 & 1(Nep) \\
\dpn{2013}{FY}{27}  &  30 & 26 &  4 & 0 \\
Orcus               &  26 & 18 &  8 & 0 \\
\dpn{2007}{UK}{126} &  12 &  8 &  4 & 0 \\
\dpn{2010}{RF}{43}  &   5 &  4 &  1 & 0 \\
\dpn{2002}{TC}{302} &   4 &  4 &  0 & 0 \\
\dpn{2003}{AZ}{84}  &   3 &  1 &  2 & 0 \\
\dpn{2002}{UX}{25}  &   1 &  1 &  0 & 0 \\
\dpn{2010}{EK}{139} &   1 &  1 &  0 & 0 \\
\hline
\thead{Total for \hfill \null \\ the sample} \hfill \null & 378 & 305 & 69 & 4
\enddata
\end{deluxetable}

Table \ref{tab:ejects} shows the list of objects ejected and the end state of their ejections. Overall 378 (out of 6800; 5.56\%) of the orbits were ejected from the solar system; this implies that the solar system is still evolving and 1 Gyr into the future will have lost $\sim$2 of these objects. Most of the ejections (305 out of 378; 80.7\%) were into outer space, with an important fraction (69 out of 378; 18.2\%) falling into the inner solar system. Finally a few of the orbits (4 out of 378; 1.1\%) end up colliding with a giant planet (2 with Jupiter and 2 with Neptune).

Encounters between DPs and giant planets are to be expected over the lifetime of the solar system. In fact, collisions of even larger objects are frequently used to explain some of the striking characteristics of the giant planets such as: the capture of Triton by Neptune \citep[e.g.][]{Agnor06}, the tipping of Uranus's rotational axis\citep[e.g.][]{Slattery92}, Jupiter's core composition \citep[e.g.][]{Liu19} and even a possible origin for Saturn's rings\citep[e.g.][]{Dubinski19}.

While the 4 collisions presented in table 4 lead to interesting possibilities, it is potentially much more interesting (and  important) the presence of 69 orbits that fall into the inner solar system. In the case of such an event, the infalling DP would become the equivalent of a huge short period comet, with a longer physical life span that would permit tens of thousands of orbits and could potentially endanger Earth as we know it. Unfortunately, we cannot follow the evolution of these 69 orbits, since the setup of our simulations does not allow us to follow the evolution of objects with orbits inside 5 au (inside Jupiter's orbit). 

These 69 orbits represent an average infalling rate of 1.01\% per DP per Gyr, and an overall 34.5\% probability when considering all 34 DPs we are tracking. This probability is slightly smaller than the rate of cometary infall from the Kuiper belt found by \citet{Munoz19}, that amounts to 4.6\% per cometary nuclei per Gyr; this difference is not entirely unexpected since lighter objects are easier to move around. However, the presence of a DP in the inner solar system, would certainly be much more spectacular. 

As far as life on Earth is concerned, any object larger than about 20 km could, potentially, become a game changer. There are expected to be approximately $10\,000$ times more objects between 20 and 200 km than those on our sample \citep{Fraser09,Dones15}, this implies an infall rate of a few thousand objects larger than 20 km each Gyr (or equivalently, a few every Myr).

Specifically, the most likely object to reach the inner solar system within the next Gyr is Haumea, while \dpn{2002}{MS}{4}, \dpn{2007}{OR}{10}, and Orcus, have a modest probability and \dpn{2015}{RR}{245}, \dpn{2013}{FY}{27},  \dpn{2007}{UK}{126}, \dpn{2003}{AZ}{84}, and \dpn{2010}{RF}{43} have a minor probability of reaching the inner solar system (see table \ref{tab:ejects}). It must be also noted that if Haumea (and to a slightly lesser extent \dpn{2007}{OR}{10}) don't reach the inner solar system within the next Gyr they become less likely to do so, as they are also very likely to be ejected from the solar system. In general a critical step for infalling orbits includes a close encounter with Neptune, but such an encounter is approximately four times more likely to result in an expulsion of the solar system than in an infalling trajectory.

The most likely fate for objects ejected from the Kuiper belt is the expulsion of the solar system. Haumea and \dpn{2007}{OR}{10} are the objects more likely to do so, with a half life of about 1 Gyr; \dpn{2002}{MS}{4}, \dpn{2015}{RR}{245}, \dpn{2013}{FY}{27}, and Orcus are also quite likely to be ejected with a half life of 2 - 4 Gyr, and \dpn{2007}{UK}{126}, \dpn{2010}{RF}{43}, \dpn{2002}{TC}{302}, \dpn{2003}{AZ}{84}, \dpn{2002}{UX}{25}, and \dpn{2010}{EK}{139}, having a non negligible probability of ejection, but with a half life longer than that of the solar system. In summary we expect about 1.5 DPs to be expelled from the solar system in the next Gyr.

\subsection{Future stability of the outer solar system}
\label{subsec:FutStab}

When trying to paint a picture of the Kuiper belt 1 Gyr into the future, the most striking feature is the overall loss of approximately 2 objects (1.89 objects lost).

Another characteristic is that there is no object for which its 200 orbits converge in the same point after 1 Gyr (even for Sedna, the slowest and most stable of the DPs, the behavior of the different runs shows we can not keep track of the mean anomaly beyond 10 Myrs). If we ignore the mean anomaly, we can only have some confidence on the argument of pericenter and longitude of the ascending node for Sedna and \dpn{2012}{VP}{113}, for all other objects we lose coherence for both quantities before the 50 Myr mark; on the other hand, our simulations do not consider the interactions with external potentials (such as stellar flybys, or the effect of Planet 9); any such encounter would affect more severely precisely those objects that are particularly well behaved in our simulations.

This does not mean that we will be completely unable to recognize some of the orbital characteristics of the trans-Neptunian DPs in the next Gyr:

Focusing only on the 6 DPs we have classified as resonant, we see that they are extremely likely to remain in the Solar System (5.96 out of 6, 99.3\%); in fact, they are likely to remain inside their current MMRs (5.915 out of 6, 98.6\%), but only slightly less than half will remain with similar values of their main orbital parameters (2.89 out of 6, 48.2\%).

Although the 28 non-resonant objects are less likely to remain in the solar system (26.15 out of 28, 93.4\%), they are more likely to maintain similar values of their main orbital parameters (20.135 out of 28, 71.9\%). 

Overall, by focusing on the main orbital parameters, we would be able to recognize about two thirds of the DPs (23.025 out of 34, 67.7\%), and about three fourths (26.05 out of 34, 76.6\%) if we allow for the resonant objects that remain inside their MMR.

\subsection{Effect of Additional Unseen Populations}

When considering together this work (which deals with DPs and TNOs greater than about 400km) and our previous work \citep[][which deals with small TNOs in the 2-10 km range]{Munoz19}, there is a gap for a population of intermediate sized objects (20-400 km). These objects are able to also have a gravitational effect on other objects, but overall we expect their disruptive effect to be much smaller than the one produced by our sample; we would expect that including 100 additional objects (the most massive within this population) would increase the disruption by less than 10\% but with an increase of a factor of $\sim$ 5 on the computational effort, while a massive simulation that includes the 1000 most massive objects of this population, would include nigh all of the disruptive effect due to the rapid decay in the mass distribution, and increase the disruption by less than 20\% but with an egregious increase on the computer expense. 

A different possibility would be the presence of additional large TNOs, not included in our sample (larger than 400 km, or even 1000 km). For example \citet{Schwamb09} and \citet{Shankman17} estimate between 40 and 80 Sedna-like objects in the outer solar system; these objects are expected to be similar in size, eccentricity, and perihelia to Sedna. Such population would have very little effect on our 34 object sample; we can see that Sedna is very stable, this implies that no other body from our sample interacts with it, equivalently Sedna does not interact with any other object; also with a perihelion of 76 au it never gets close enough to perturb objects that could potentially get close to Neptune to be strongly disrupted. Also, most of the effect between DPs is secular in nature, and any object with such orbital characteristics would spend too little time close to other objects for the secular interactions to be significant. Interactions between pairs of Sedna-like objects would also be unimportant due to the large volume at such large distances from the Sun.

A potentially more disruptive population would be objects with semimajor axis similar to Sedna, but with perihelia able to reach the Classical Kuiper Belt region; any such object would be much less efficient than the DPs from our sample since they spend only a very small fraction of their time within the Kuiper belt range. Since only about 1\% of their time would be spent in the Classical Kuiper Belt region, approximately 100 such objects would be needed to contribute as much as any single object from our sample, and a few hundred would be needed to change our overall conclusions. While objects with this characteristics can not be ruled out, such a large population is not expected to be hidden from us. 

While the existence of additional large and distant DPs would decrease the stability of our sample, and our results thus represent a lower limit to the rate at which DPs will be lost, we do not expect there are enough additional interactions to change our conclusions significantly.






\subsection{Other Considerations}


This study explores the future evolution of the trans-Neptunian region. It would be interesting to know how this region has evolved to reach its present configuration. It is obvious the number of DPs is falling with time, this also means that the evolution speed is slowing down with time. 

According to \citet{Munoz17} the evaporation rate is proportional to the square of the mass of the disk, in systems with an interior giant planet; however such models ignore the possibility of additional perturbations, such as flybys or giant planet migration. However, while it is possible to do a statistical study of the evolution tempo, it is not possible to do a detailed study of the number of objects that were lost in the previous Gyr, since the detailed efficiency depends on the detailed configuration of a past forgotten time (be it 1 Gyr into the past, or at the beginning of the solar system).

While we have made use of complex simulations to explore the future evolution of the outer solar system, it is arguably worth considering if a simpler (and faster) model is able to provide us with the conclusions about the future stability of the system reached so far. This can be accomplished by treating our DP sample as mass-less test particles, recognizing that the orbital dynamics are dominated by the giant planets.

We ran 200 additional simulations with identical parameters and initial conditions as those of Section \ref{ssec:dynmodel}, but considering only the gravitational perturbations of the four giant planets over the 34 DPs, which were treated as test particles. Each simulation differs from the others by the initial time step of the integrator, varied from 350 to 450 days on 0.5 day intervals. The simplified simulations were approximately 4 times faster that our full N-body simulations. 

An analysis over the 200 simplified simulations shows results that are in general equivalent to the full N-body runs. There were a total 2221 irregular orbits and 372 ejections in the simplified runs, as opposed to 2195 irregular orbits and 378 ejections in the full runs. Although, globally, these results are interchangeable, when looking at the behaviour of individual objects there are some irreconcilable differences.

First, out of the 22 objects that have at least 1 irregular orbit in table 3, 10 present statistically different results within the simplified simulations. When comparing the full runs to the simplified runs we find that 3 DPs are more stable, 6 DPs are less stable, and 1 DP has more irregular orbits but less ejections. Two objects in particular stand out, namely Haumea and Orcus. For Haumea, the simplified runs produce 146 ejections versus 125 in the full runs; for Orcus, not a single ejection is obtained in the simplified runs, additionally it remains well locked inside the 3:2 MMR in all cases; in comparison the full runs produce 26 ejections and only 161 regular orbits in $a$. The latter result highlights the importance of self interactions between massive DPs, particularly inside MMRs, where mutual distances between the objects can be significantly reduced and secular perturbations accumulate more efficiently. This shows that the effect of DPs mutual perturbations on the secular evolution of the Kuiper belt cannot be in general ignored, otherwise some of the long-term conclusions will be flawed.

Finally, besides the statistical analysis of the whole sample, we consider that the evident chaotic nature of some of the DPs (e.g. Haumea, \dpn{2007}{OR}{10}, or \dpn{2015}{RR}{245}) calls for an individualized analysis of their expected evolution; we intend to pursue such analysis in a future paper (Mu\~noz-Guti\'errez et al. in preparation).

\section{Summary and Conclusions}

In this work we have explored the global stability of a vast region of phase-space in the outer solar system, where the majority of the largest known TNOs are located. We have focused most of our work on the 34 largest known TNOs, which we call indistinctly DPs.

We performed a frequency analysis, considering the simplest model of the solar system (i.e. Sun + giant planets), running thousands of orbits in a region broadly covering from 30 to 82 au in $a$, and from 0 to 0.6 in $e$. 

The frequency map analysis (FMA) method allows us to estimate with some certainty the stability of the outer solar system, using only very short numerical integrations. Most of the 34 DPs in our sample were found to be located in stable areas of the produced diffusion maps, both when considering $0^\circ$ inclination or a random inclination between $0^\circ$ and $50^\circ$.

The random inclination simulations also allowed us to determine, by means of the diffusion time, the stability of Kuiper belt objects as a function of their perihelion. We found, as can be expected, that the greater the perihelion, the longer the stability time.

Overall, the large diffusion maps we used to study statistical stability are convenient for the analysis of comets (or any large set of objects) but, for our list of 34 specific DPs, a custom made map for each object will give us a better perspective on the evolution of each of them. The analysis of the individual maps suggests that only 5 objects from our sample are unstable on timescales of 1Gyr.

While a FMA will provide a certain degree of confidence for the evolution of any given particle, these are lacking due the omission of the DPs. This omission impacts the results in two relevant areas: firstly, DPs are the second most important ingredient on the stability of the Kuiper belt region and, secondly, the effect of the DPs on the stability is slower to take hold than the one from the giant planets, thus FMAs are not entirely suited to predict the long-term behaviour in every region of phase-space.

A better strategy to study the stability of these DPs consists on running a more complete model of the solar system and to integrate it for a longer period of time. 

Since the Kuiper belt region of the full model is chaotic, a statistical analysis is necessary; to do this, we decided to run an ensemble of models with identical initial conditions but random perturbations (arising from the time steps of the numerical integrator). We performed 200 long-term simulations considering the mutual perturbations of the giant planets and the 34 largest TNOs; our final analysis consists of 6800, 1 Gyr long, individual realizations (200 for each of the 34 DPs).
 
From our simulations we divide the 34 DPs in three categories based on the long-term behavior of their 200 realizations. We find that 6 DPs are ``Resonant'' (where most of their realizations remain within a known resonance); of the 28 non-resonant objects 17 are ``Stable'' (where most of their realizations present very small evolution on the $a$, $e$, and $i$) while 11 are ``Unstable'' (where many of their realizations change drastically in at least one of their main orbital parameters).

We also find that 12 DPs present at least one realization where they are unable to finish the simulation; either by being ejected from the solar system, by colliding with a giant planet, or by falling to the inner solar system. Out of these 12 objects we classified 10 as Unstable and 2 as Resonant.

When studying the global evolution of the Kuiper belt over the next Gyr, we find that statistically: 23.025 DPs are expected to remain with recognizable orbital parameters, 9.085 are expected to have drastic transformations in at least one of their main orbital parameters, and 1.89 are expected to be lost altogether.

Of the 1.89 lost objects: 1.525 DPs are expected to leave the solar system altogether, 0.345 are expected to find their way to the inner solar system (i.e. to reach distances of less than 5 au from the Sun), and 0.02 are expected to collide with a giant planet.

An unexpected result from this work is that objects like Haumea, \dpn{2007}{OR}{10}, and \dpn{2015}{RR}{245}, turn out to be in highly unstable orbits. Though it may be thought that these DPs are permanent fixtures of the solar system, in reality it is highly unlikely that all three of these objects will be part of the solar system inventory 1 Gyr into the future. While interesting, we are leaving a detailed analysis of the chaotic nature of these and other DPs in our sample for a future paper.

\begin{acknowledgments}

We thank the anonymous referee for an insightful report, which helped to improve the quality of this paper. AP would like to thank grant PAPIIT IG 100319.

\end{acknowledgments}

\bibliography{dpsbib}{}
\bibliographystyle{aasjournal}


\end{document}